\documentclass{article}
\usepackage{amssymb}
\usepackage{amsmath}
\usepackage{enumerate}
\usepackage{graphicx}
\usepackage[english]{babel}
\usepackage{multirow}
\usepackage{booktabs}
\usepackage{array}
\usepackage{todonotes}
\newcolumntype{L}[1]{>{\raggedright\let\newline\\\arraybackslash\hspace{0pt}}m{#1}}
\newcolumntype{C}[1]{>{\centering\let\newline\\\arraybackslash\hspace{0pt}}m{#1}}
\newcolumntype{R}[1]{>{\raggedleft\let\newline\\\arraybackslash\hspace{0pt}}m{#1}}

\begin{document}

\begin{Large}

A novel deep learning-based method for prostate segmentation in T2-weighted magnetic resonance imaging

\end{Large}

\vspace{5mm}

Davood Karimi, Golnoosh Samei, Yanan Shao, Septimiu Salcudean 

\vspace{5mm}

Department of Electrical and Computer Engineering, The University of British Columbia, Vancouver, BC, Canada

\begin{abstract}

We propose a novel automatic method for accurate segmentation of the prostate in T2-weighted magnetic resonance imaging (MRI). Our method is based on convolutional neural networks (CNNs). Because of the large variability in the shape, size, and appearance of the prostate and the scarcity of annotated training data, we suggest training two separate CNNs. A global CNN will determine a prostate bounding box, which is then resampled and sent to a local CNN for accurate delineation of the prostate boundary. This way, the local CNN can effectively learn to segment the fine details that distinguish the prostate from the surrounding tissue using the small amount of available training data. To fully exploit the training data, we synthesize additional data by deforming the training images and segmentations using a learned shape model. We apply the proposed method on the PROMISE12 challenge dataset and achieve state  of the art results. Our proposed method generates accurate, smooth, and artifact-free segmentations. On the test images, we achieve an average Dice score of 90.6 with a small standard deviation of 2.2, which is superior to all previous methods. Our two-step segmentation approach and data augmentation strategy may be highly effective in segmentation of other organs from small amounts of annotated medical images.

\end{abstract}

\section{Introduction}

Segmentation of the prostate in T2-weighted magentic resonance imaging (MRI) is an essential step for many tasks in treatment planning and intervention \cite{leslie2012,zini2012}. Automatic segmentation methods are highly desirable because they can increase the speed and reproducibility of the segmentation. In the past decades, many studies have proposed (semi-)automatic methods for prostate segmentation in T2-weighted MRI \cite{zhu2006,ghose2012,litjens2014b}. However, fully automatic prostate segmentation is very challenging because of the inter-patient variability in the prostate size, shape, and appearance, variations in the scanners and scanning protocols, and similarity of the prostate with the surrounding tissue.

A large number of the methods proposed for prostate segmentation in T2-weighted MRI use atlases \cite{klein2008,martin2010}. In these methods, a number of MR images with known prostate segmentation are registered to the target image. Mutual information, cross-correlation, image feature correspondence, and the image gradient are among the image similarity metrics used for registration. The deformed prostate segmentation masks of the atlas images are then combined to infer the segmentation of the prostate in the target image. Therefore, atlas-based methods turn the segmentation problem into a registration problem. A critical choice in these methods is how to combine/fuse the registered segmentation masks. One can rank the segmentation masks based on some image similarity metric and choose the most similar segmentation, or use more elaborate methods such as majority voting, simultaneous truth and performance level estimation (STAPLE), or iterative label fusion \cite{dowling2011,klein2008,langerak2010}. In general, atlas-based methods are computationally expensive and can produce poor segmentations, especially if the target image is very different from the population of the images in the atlas. To achieve acceptable results, some atlas-based methods rely on additional steps based on statistical shape models  \cite{martin2010,martin2008,gao2010}. Moreover, most of the atlas-based methods follow a global registration strategy, which makes them unnecessarily sensitive to the anatomical features that are far away from the prostate and increases the computational time. To overcome these shortcomings, some studies have proposed two-step registration approaches in which a global registration is first performed to identify the location of the prostate in the image. In the second stage, a local registrtation is performed by focusing on the prostate region \cite{tian2015,ou2012}.

Another class of methods includes those based on deformable models such as active shape models and level sets \cite{zhu2007,liu2009,kirschner2012,toth2012,vincent2012}. A great appeal of these methods is that they are based on sound theory from physical sciences and mathematics. However, these methods can be very sensitive to initialization \cite{toth2011} and a good initialization may be hard to obtain. Moreover, the quality of segmentation can be poor, especially where the edge information is not strong. Therefore, some of these methods depend on manual initialization or rely on other prioir information in the form of shape models to regularize or refine the generated segmentation \cite{yang2017,cootes1993}.

Some studies have proposed methods based on graph cuts \cite{egger2013,mahapatra2014}. Although these methods are versatile, they have their own limitations. For example, they produce poor results at the locations of weak edges and typically need post-processing steps in order to obtain satisfactory results. Recently, some studies have shown that the performance of graph cut-based methods can be substantially improved by using active contours and by formulating the graph cut method in terms of super-voxels instead of raw voxel intensities \cite{tian2016,tian2017}.

Because of the difficulties faced by the methods mentioned above, a large number of studies have tried to combine the advantages of two or more of these frameworks. Many of these methods also use some type of machine learning to achieve improved results. For example, several studies have combined probabilistic learning of the distribution of prostate texture or voxel intensities with shape models \cite{toth2011b,makni2008,allen2006}. Supervised and un-supervised machine learning methods such as random forests and clustering methods have also been combined with deformable models and atlas-based methods for prostate sementation in T2-weighted MRI \cite{makni2014,ghose2012b,gao2014}.  One study has suggested using shape models in the framework of marginal space learning for prostate segmentation in T2-weighted MRI \cite{birkbeck2012}

Despite the great efforts and numerous methods that have been proposed in recent years, automatic segmentation of the prostate in T2-weighted MRI still remains a challenge. Most of the proposed methods achieve much lower performance than manual segmentation. If the test images are very different from the images used for model development, e.g., due to inter-patient variability or different scanning protocols, the performance of these methods can deteriorate substantially. 

In recent years, deep convolutional neural networks (CNNs) have achieved unprecedented results in segmentation of natural images \cite{long2015,chen2016,noh2015}. Compared to the more traditional segmentation methods, the new CNN architectures that have been proposed for dense segmentation possess a number of highly desirable characteristics: 1) they have a very high capacity that enables them to effectively describe the large variations that exist in the training data, 2) they are able to explain local and global information at different resolutions simultaneously, 3) in many applications they can achieve quite satisfactory results without the need to additional post-processing steps to refine their segmentation, which also implies that they can be trained end-to-end as a single module, and 4) even though they have long training times, their inference time is very fast. Consequently, many studies have recently employed CNNs for segmentation of medical images \cite{litjens2017} and, in general, they have reported very promising results. For segmentation of the prostate in T2-weighted MRI, in particular, deep CNNs with volumetric convolutional filters have been shown to achieve very good results \cite{milletari2016}. One study resampled the ground-truth segmentation to generate prostate masks with different resolutions for more effective training of a deep CNN \cite{yu2017}. The trained CNN was applied on sub-volumes of the input image and averaging of the probability maps estimated for all sub-volumes was used to obtain the final prostate segmentation. The proposed method achieved state of the art results, which was also attributed to the use of short and long residual connections in the network. Another study proposed a deep CNN with 2D and 3D residual connections and achieved state of the art results \cite{sciolla2017}. Karimi et al. proposed to segment the prostate in MR images using a novel network architecture that predicted the location, orientation, and the coefficients of a statistical shape model \cite{karimi2018prostate}.

In this paper, we propose a new CNN-based method for segmentation of the prostate in T2-weighted MR images. We argue that the difficulty in achieving human-level performance in this task is due to the large variability in the shape, size, and appearance of the prostate in these images. Based on the results achieved by CNNs in segmentation of natural images, we think that theoretically they should be able to achieve human-level performance in prostate segmentation in T2-weighted MRI. However, this is not easy to achieve in practice because it is hard to effectively train large CNNs with small amounts of annotated data. To reduce this gap and effectively utilize the capacity of deep CNNs with limited training data, we suggest two strategies:

\begin{enumerate}

\item We suggest training two CNNs. The first, \textit{global}, CNN will accept the entire image as its input and generate a soft prostate segmentation mask. This initial segmentation is then used to determine the location and the extent of the prostate in the image. A second, \textit{local}, CNN will then work on a sub-volume of the image that is resampled from the input image in such a way that the prostate is approximately at the center of the volume and has an approximately fixed size. This will allow the local CNN to focus on learning features that are most relevant for accurate delineation of the prostate boundary, which is a major challenge due to similarity with the surrounding tissue, large variability, and scarcity of training data.

\item We use massive data augmentation for training of the two CNNs. Here, our argument is that even 50 training images are not sufficient to train large CNNs. Therefore, we synthesize additional realistic data by deforming the training images and their segmentation masks using displacement fields that are computed based on a prostate shape model. To further improve the training and avoid local minima, random displacements and noise are introduced during training. Moreover, using cross-validation we will identify the images that are most difficult to segment and will use this information in training our final model.

\end{enumerate}

\section{Materials and methods}

\begin{figure}[!hb]
    \centering
    \includegraphics[width=0.8\textwidth]{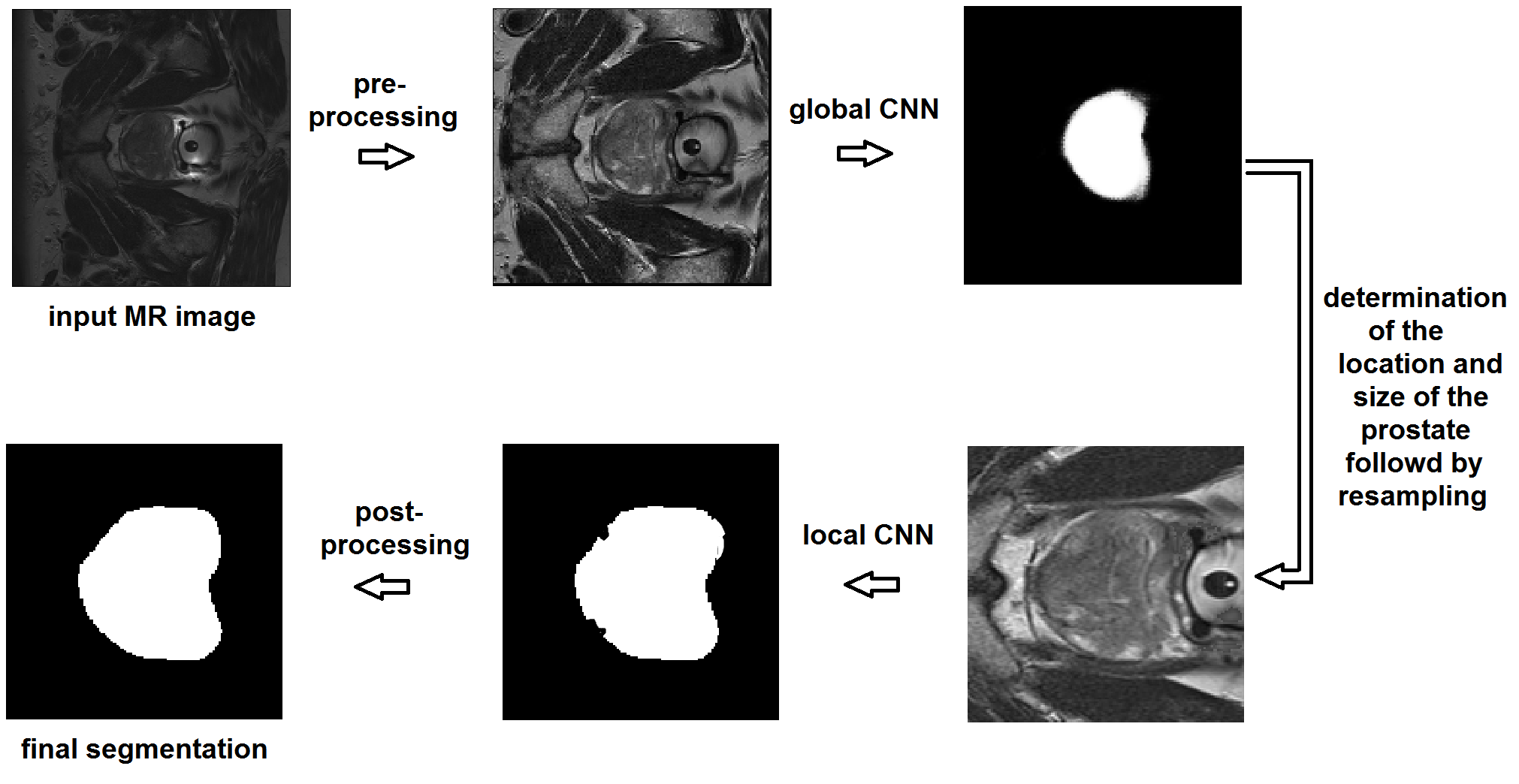}
    \caption{A schematic representation of the steps involved in the proposed segmentation method.}
    \label{fig:method_schematic}
\end{figure}

A schematic representation of the steps involved in our fully-automatic segmentation method is shown in Figure \ref{fig:method_schematic}. Considering the great variability in the shape, size, texture, and appearance of the prostate and its surrounding organs and the limited availability of training data, we suggest learning two separate CNNs. After some pre-processing steps, the input image is passed to a global CNN, which generates an initial soft segmentation of the prostate. We will use this initial segmentation, along with a prostate shape model, to identify the location and the extent of the prostate in the image. Using this information, we extract a volume of interest that contains the prostate and send it to a second, local, CNN that accurately delineates the prostate boundary. Basic thresholding and post-processing steps on the output probability map of the local CNN will generate the final prostate segmentation mask. In the following sub-sections, we first provide a description of the steps involved in our proposed segmentation method. We will then provide more detailed information on various steps.

\subsection{Steps of the proposed segmentation method}

As shown in Figure \ref{fig:method_schematic}, given a test image the following steps are performed to generate the prostate segmentation mask.

\begin{description}

\item[Pre-processing] First, a bias correction is applied to the image using the N4ITK algorithm \cite{sled1998,tustison2010}. 
The image is resampled to an isotropic voxel size of $1 \text{ mm}^3$. A volume of size $128 \times 128 \times 72$ voxels is then created, using either centered-cropping or zero-padding if the image is, respectively, larger or smaller than this desired size in each dimension. Finally, the image is normalized such that the voxels have a mean of zero and a standard deviation of one.

\item[Determining the prostate location and extent] The pre-processed $128 \times 128 \times 72$-voxel image is sent to the global CNN. The output of this network is a probability map where the value of each voxel shows the probability of that voxel belonging to the prostate. As we will show in the Results section, this probability map itself provides a good segmentation of the prostate. However, in our method we will treat it as an initial segmentation that is used only to identify the location and the extent of the prostate in the image. The center of mass of this segmentation can give us an accurate estimation of the location of the center of the prostate. However, in order to obtain an accurate estimation of the extent of the prostate, we need to remove the possible outlier voxels. For this purpose, we fit a prostate shape model to the learned probability map using a particle swarm optimization algorithm.

\item[Fine segmentation of the prostate] Given the information on the location and the extent of the prostate from the above step, the pre-processed MR image is resampled again to generate an input for the local CNN. Similar to the global CNN, the local CNN has an input size of $128 \times 128 \times 72$ voxels. However, whereas the input to the global CNN has an isotropic voxel size of $1 \text{ mm}^3$, the input to the local CNN is generated such that the prostate is at the center and has a pre-determined size. Specifically, in this stage we resample the image such that the prostate has a size of approximately $80 \times 80 \times 48$ voxels. For example, if the prostate size estimated using the global CNN is $40 \times 32 \times 36 \text{ mm}$, the MR image is resampled to have a voxel size of $0.50 \times 0.40 \times 0.75 \text{ mm}$ such that the prostate in the resampled image has, approximately, the desired size of $80 \times 80 \times 48$ voxels. This resampled volume of interest is passed to the local CNN, which estimates a more accurate segmentation of the prostate.

\item[Post-processing] The probability map produced by the local CNN, thresholded at 0.50, constitutes our penultimate prostate segmentation. Our post-processing consists only of applying an opening operation (erosion followed by dilation) \cite{serra1986} using a spherical structuring element with a radius of $2 \text{ mm}$. This operation will remove the occasional non-smooth artifacts on the surface of the prostate mask obtained by thresholding the probability map.

\end{description}

\subsection{CNN architecture}

\begin{figure}[h]
    \centering
    \includegraphics[width=\textwidth]{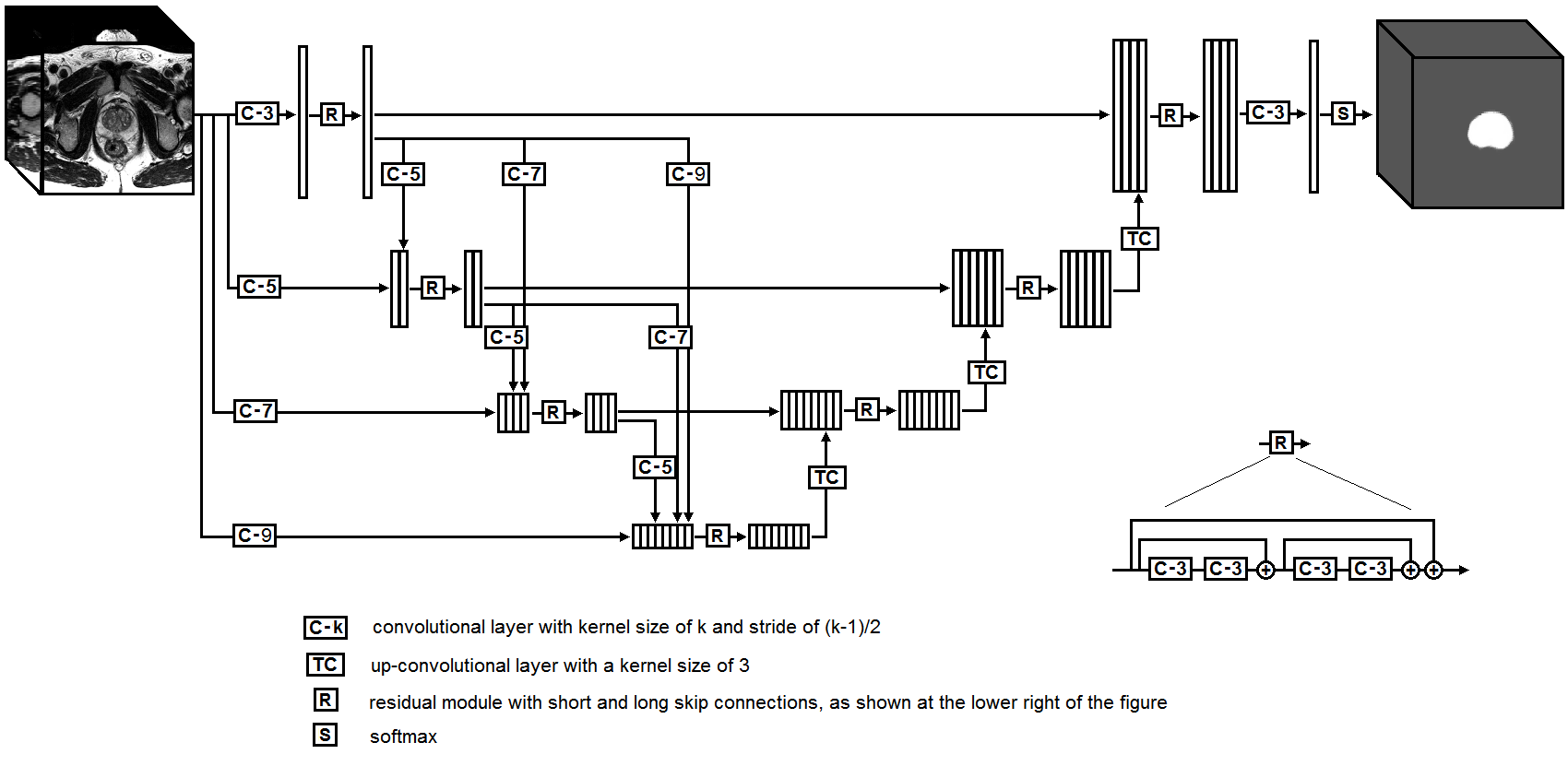}
    \caption{A schematic representation of the CNN architecture used in this study. The network shown in this figure has a depth of 4. The global and local networks used in this study had depths of 5 and 3, respectively.}
    \label{fig:cnn_schematic}
\end{figure}

The CNN architecture used in this study takes advantage of many of the effective CNN design practices. A schematic representation of this architecture is shown in Figure \ref{fig:cnn_schematic}. The network has a contracting and an expanding path that enable it to learn features at fine and coarse resolutions \cite{ronneberger2015,cciccek2016}. In the contracting path, the network computes convolutional feature maps with kernels of increasing size $k \in \{ 3, 5, ... , 2d+1 \}$ and stride $s \in \{ 1, 2, ..., d \}$. The network shown in Figure \ref{fig:cnn_schematic} is for $d= 4$. These convolutional feature maps will capture different levels of detail directly in the source image. In each resolution level, a residual module with short and long skip connections is used to increase the expressiveness of these features and network's capacity and also make the training faster \cite{he2016,drozdzal2016}. Each of the computed feature maps is passed forward to all lower levels by applying convolutional filters with the proper stride. This will promote feature reuse and make the training easier by reducing the number of parameters and improving the gradient flow \cite{huang2016}. Therefore, in the contracting path the number of feature maps geometrically increases while their size becomes smaller. The computed features then go through a series of up-convolutions in order to build up the prostate segmentation mask. Thsi process will start from the low-resolution feature maps of the contracting path, which moslty contain coarse information. In order to aid recovery of the fine detail, the multi-resolution information available in the feature maps of the contracting path is re-introduced into the expanding path via concatenating those feature maps to the feature maps of the expanding path. Residual modules, similar to those in the contracting path, are also applied here in the expanding path. All convolutional and up-convolutional layers, except for those applied directly on the input MR image, use kernels of size 3. Furthermore, all these layers, except for the last are followed by rectified linear units \cite{krizhevsky2012}. The final feature map, which has the same dimensions as the source image is passed to a final convolutional layer followed by a soft-max operation to produce a pixel-wise probability map of the prostate.

Both CNNs follow the same architecture as shown in Figure \ref{fig:cnn_schematic}, except that they have different depths. We found that for the global CNN the results were improved as we increased the network depth to $d=5$. We think this is because the low-resolution feature maps have a larger field of view that enables them to learn features that better represent the location and extent of the prostate in the image. For the local CNN, however, a shallower network with a depth $d=3$ led to better results. This is because the input to this network is a resampled volume of interest in which the prostate is approximately at the center and has approximately a fixed pre-determined size. Therefore, the local CNN does not have to learn to estimate the location or the extent of the prostate but only to delineate the prostate boundary. Therefore, for the local CNN increasing the depth unnecessarily increases the number of network parameters and introduces coarse features that do not contribute substantially to the delineation of the prostate boundary.

\subsection{Training and evaluation}

We used the data from the PROMISE12 challenge \cite{litjens2014b}. This dataset consists of 50 training and 30 test images, which have been acquired at different centers and using different scanners and scanning protocols. The dataset is very challenging because of the large variation in the voxel size, field of view, and dynamic range of the images as well as in the appearance of the prostate. Approximately half of the images include an endorectal coil.

We used a five-fold cross validation approach to determine the network size, number of features and other important training parameters such as the learning rate scheduling and the dropout rate. Each time, the two CNNs were trained on 40 of the images and evaluated on the remaining 10 images. Having found the optimal network size and learning parameters using cross-validation, the final global and local CNNs were trained on all 50 training images and applied on the 30 test images. The final global and local CNNs were both initialized at random using He's method \cite{he2015} and trained using Adam's method \cite{kingma2014} with a batch size of one to maximize the Dice score between the output probability map of the network and the ground-truth segmentation mask. Training was performed for 1000 epochs. The starting learning rate was $10^{-5}$, which was multiplied by 0.5 when the training cost function plateaued.

In order to cope with the limited training data, we use a data augmentation scheme based on a learned shape model. In this strategy, the training images and its labels are deformed using displacements suggested by a learned shape model. Moreover, we add white Gaussian noise with a standard deviation of 0.03 to each image at each training step. To further reduce the risk of overfitting, we use dropout \cite{srivastava2014} with a rate of 0.15 on all convolutional layers in the network. This dropout rate significantly reduced the overfitting such that the segmentation performance on the training and validation images was almost equal.

\section{Results and discussion}

Table \ref{table:dice_table} shows the average and standard deviation of the Dice score on the training, validation, and test images. In order to show the effect of the different steps in our segmentation pipeline, we have shown the resulting Dice score after each step. Our proposed method achieves a high final Dice score of $90.6$ with a low standard deviation of $2.2$.

\begin{table}[!htb]
\footnotesize
  \begin{center}
    \begin{tabular}{  l C{2.75cm} C{2.75cm} C{2.75cm}}
\hline
 &  Global CNN & Local CNN & Post-processing \\ \hline
Training     & $85.0 \pm 3.9$ & $91.0 \pm 2.3$ & $91.2 \pm 2.2$  \\ 
Validation   & $84.9 \pm 4.1$ & $90.4 \pm 2.3$ & $91.2 \pm 2.0$  \\
Test         & $85.4 \pm 3.6$ & $90.2 \pm 2.4$ & $90.6 \pm 2.2$  \\  \hline
    \end{tabular}
  \end{center}
  \caption{\footnotesize{Mean$\pm$standard deviation of the Dice score for training, validation, and test images after each step in the segmentation pipeline.}}
  \label{table:dice_table}
\end{table}

\begin{figure}[h]
    \centering
    \includegraphics[width=0.5\textwidth]{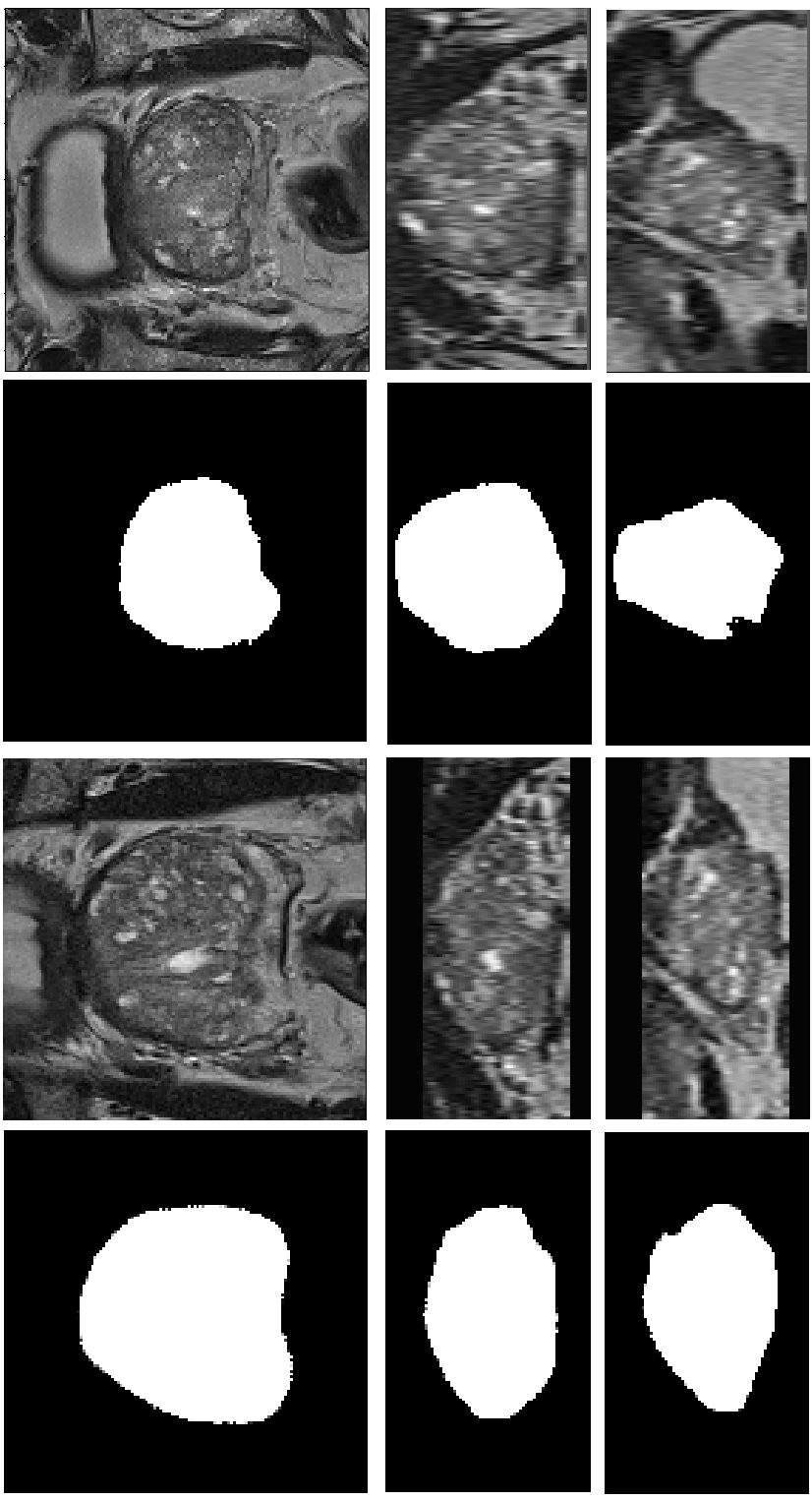}
    \caption{The prostate segmentation masks produced by the global and local CNNs. The first row shows the $128 \times 128 \times 72$-voxel image sent to the global CNN. The second row shows the segmentation mask produced by the global CNN. The third row shows the volume resampled from the input image to be used by the local CNN. The last row shows the prostate segmentation mask produced by the local CNN.}
    \label{fig:sample_segmentation}
\end{figure}

A test image and its prostate segmentation masks produced by the global and local CNNs in axial, sagittal, and coronal directions are shown in Figure \ref{fig:sample_segmentation}. The accuracy of segmentation of the global CNN is not very high. The Dice score achieved by the global CNN is $85.4 \pm 3.6$, which is substantially lower than that achieved by the local CNN ($90.2 \pm 2.4$). Nonetheless, as we mentioned above, our method uses the segmentation provided by the global CNN only to estimate the location and extent of the prostate in the image. Our results show that it is possible to accurately extract this information from the probability map produced by the global CNN. The error in the estimation of the start and end of the prostate in axial, sagittal, and coronal directions were, respectively, $0.8 \pm 0.9$, $1.4 \pm 1.5$, and $1.2 \pm 1.0$ millimeters, and the error in estimating the size of the prostate in these three directions were, $1.6 \pm 1.4$, $2.2 \pm 2.0$, and $2.1 \pm 1.7$ millimeters, respectively. These errors are quite within the range of acceptable error range for our intended goal. Using the information on the size and location of the prostate in the image, a $128 \times 128 \times 72$-voxel volume is resampled such that the prostate is approximately at the center and has a size of $80 \times 80 \times 48$ voxels. As can be seen in the example image shown in Figure \ref{fig:sample_segmentation}, depending on the size of the MR image and the size and location of the prostate, this may lead to stretching or shrinking of the image in different directions and may need cropping or zero padding. However, it will ensure that the prostate is at the center of the volume and has a pre-determined size, allowing the local CNN to achieve superior segmentation accuracy. As can be seen in Figure \ref{fig:sample_segmentation}, the local CNN generates an accurate segmentation mask that usually does not need any post-processing. However, in our method we always perform a morphological opening operation that removes any small non-smooth features caused by hard-thresholding the output probability map of the local CNN. As shown in Table \ref{table:dice_table}, this operation has a noticeable positive effect on the average Dice coefficient.

\begin{figure}[h]
    \centering
    \includegraphics[width=0.3\textwidth]{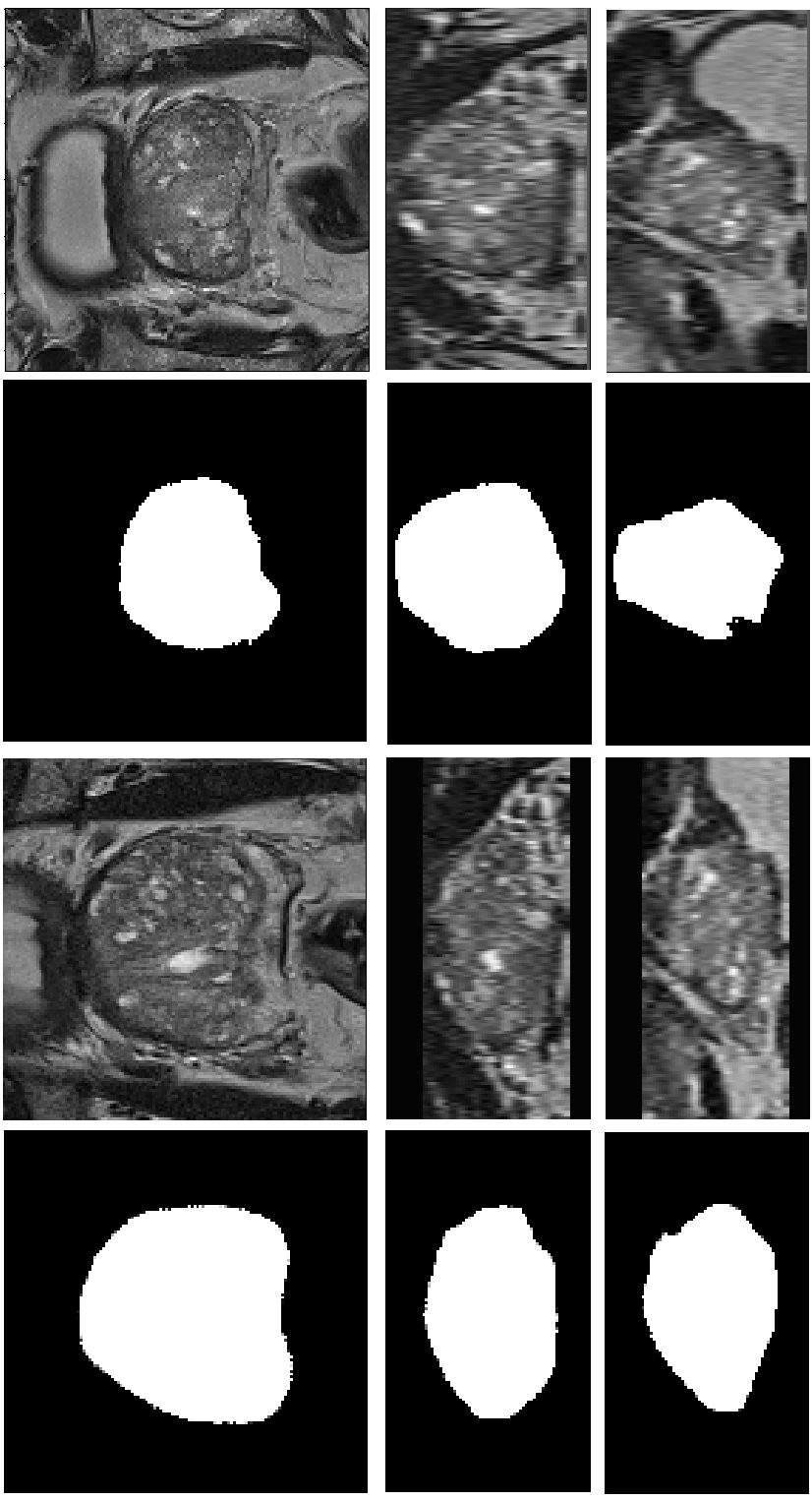}
    \caption{The prostate segmentation masks produced by the global and local CNNs. The first row shows the $128 \times 128 \times 72$-voxel image sent to the global CNN. The second row shows the segmentation mask produced by the global CNN. The third row shows the volume resampled from the input image to be used by the local CNN. The last row shows the prostate segmentation mask produced by the local CNN.}
    \label{fig:more_sample_segmentations}
\end{figure}

More examples of the performance of the proposed segmentation algorithm on test images are shown in Figure \ref{fig:more_sample_segmentations}. They further demonstrate that the proposed method is able to accurately segment the prostate. As the numbers in Table \ref{table:dice_table} suggest, one of the reasons for the success of the proposed method is the two-step segmentation method that enables the local CNN to achieve superior results. As we argued above, the variability in the shape, size, and appearance of the prostate and it surrounding anatomy is very high. Although large CNNs have the capacity to capture such complexity, the amount of training data is not large enough to realize their full potential. Therefore, in the absence of very large datasets as those available for natural images \cite{russakovsky2015}, it will be very difficult to achieve segmentation accuracies close to human experts. Our strategy of training two separate CNNs eases the situation by introducing a division of labor in which the global CNN is only expected to produce a segmentation that is accurate enough to estimate the location and size of the prostate while the local CNN is given a well-formatted input that makes it easy to accurately segment the prostate boundary.

Another important factor that contributes to the performance of our proposed method is data augmentation. In order to examine the impact of data augmentation, we trained the CNNs without any data augmentation and achieved a final Dice score of $86.1 \pm 6.6$. Another widely used data augmentation strategy is to deform the image and its segmentation mask using a random deformation field. We found that this strategy was also very useful in pushing the performance of our proposed method, but achieved a slightly lower Dice score $0.893 \pm 0.040$ that our augmentation approach based on shape models.

Another important factor in our optimization is the careful choice of the dropout rate. Dropout is known to be essential in networks with large fully-connected layers. However, its importance is usually assumed to be less significant in the convolutional layers \cite{srivastava2014}. However, our experiments showed that, especially for achieving superior performance in the local CNN, it was necessary to choose a dropout rate of $10-20 \%$. This ensured that training and validation performance were very close and the network converged nicely, whereas without dropout our method achieved a good performance on the training set but

\section{Conclusion}

Deep learning models have been very successful for medical image segmentation. Given the small size of typical datasets, training of these models is challenging. In this work, we showed that for prostate segmentation in MRI, an effective approach is to perform the segmentation in two stages: in Stage One the prostate is located and roughly segmented; in Stage Two, the prostate is accurately delineated using the bounding box obtained from Stage One.

A similar procedure as that proposed in this paper for prostate segmentation in MRI may be useful for other medical image segmentation applications.

\bibliographystyle{plain}
\bibliography{davoodreferences}

\end{document}